\documentclass[aps,prd,eqsecnum,nofootinbib,showpacs,twocolumn]{revtex4-2}

\usepackage{amsmath,amssymb,bm}
\usepackage[dvipdfmx]{graphicx}
\usepackage{color}

\numberwithin{equation}{section}

\newcommand{\be}{\begin{equation}}
\newcommand{\ee}{\end{equation}}
\newcommand{\bea}{\begin{eqnarray}}
\newcommand{\eea}{\end{eqnarray}}

\newcommand{\eq}[1]{Eq.~(\ref{eq:#1})}
\newcommand{\sect}[1]{Sec.~\ref{sec:#1}}
\newcommand{\appen}[1]{Appendix~\ref{sec:#1}}
\newcommand{\fig}[1]{Fig.~\ref{fig:#1}}
\newcommand{\tabl}[1]{Table~\ref{table:#1}}

\newcommand{\del}{\partial}

\newcommand{\eg}{{\it e.g.}}
\newcommand{\ie}{{\it i.e.}}

\newcommand{\bh}{black hole\ }

\bmdefine{\bmq}{{\bm{q}}}
\bmdefine{\bmk}{{\bm{k}}}
\bmdefine{\bmx}{{\bm{x}}}
\bmdefine{\bmy}{{\bm{y}}}
\bmdefine{\bmr}{{\bm{r}}}
\bmdefine{\bmnabla}{{\bm{\nabla}}}
\bmdefine{\bmA}{ \bm{A} }
\bmdefine{\bmD}{ \bm{D} }

\bmdefine{\bmPhi}{ \bm{\Phi} }
\bmdefine{\bmPsi}{ \bm{\Psi} }
\bmdefine{\bmcalO}{ \bm{\mathcal{O}} }

\newcommand{\nq}{\mathfrak{q}}

\newcommand{\nw}{\mathfrak{w}}

\bmdefine{\bmg}{{\bm{g}}}
\bmdefine{\bmR}{{\bm{R}}}

\newcommand{\mfh}{\mathfrak{h}}

\newcommand{\mfq}{\nq}

\newcommand{\mfA}{\mathfrak{A}}

\newcommand{\nk}{\mathfrak{p}}
\newcommand{\ki}{p}
\newcommand{\tilZ}{\tilde{Z}}

\newcommand{\hatt}{\hat{t}}
\newcommand{\hz}{\hat{z}}

\newcommand{\Za}{\tilZ}
\newcommand{\Zone}{Z}

\newcommand{\calF}{{\cal F}}

\newcommand{\tilA}{\tilde{A}}

\begin{document}


\title{Pole-skipping as missing states}
\author{Makoto Natsuume}
\email{makoto.natsuume@kek.jp}
\altaffiliation[Also at]{
Graduate Institute for Advanced Studies, SOKENDAI, 
1-1 Oho, Tsukuba, Ibaraki, 305-0801, Japan;
 Department of Physics Engineering, Mie University, 
 Tsu, 514-8507, Japan.}
\affiliation{KEK Theory Center, Institute of Particle and Nuclear Studies, 
High Energy Accelerator Research Organization,
Tsukuba, Ibaraki, 305-0801, Japan}
\author{Takashi Okamura}
\email{tokamura@kwansei.ac.jp}
\affiliation{Department of Physics and Astronomy, Kwansei Gakuin University,
Sanda, Hyogo, 669-1330, Japan}
\date{\today}
\begin{abstract}
It remains unclear in general how the pole-skipping appears as a physical phenomenon, and we study the issue in the context of the AdS soliton.
The pole-skipping has been discussed in black hole backgrounds, but the pole-skipping occurs even in the AdS soliton background. The geometry has a compact $S^1$-direction, and we compute the mass spectrum for the bulk scalar field, the bulk Maxwell field, and the gravitational perturbations with $S^1$ momentum. We show that the pole-skipping leaves its fingerprint in the the normal mode spectrum. 
The spectrum has some puzzling features because the would-be states are missing at pole-skipping points. The puzzling features disappear once one takes into account these pole-skipping points 
that we call ``missing states."
\end{abstract}
%

\maketitle

\section{Introduction}

The pole-skipping is a new universal property of Green's functions \cite{Grozdanov:2017ajz,Blake:2018leo,Grozdanov:2019uhi,Blake:2019otz,Natsuume:2019xcy}.%
\footnote{Since then, various aspects of the pole-skipping have been investigated (see, \eg, Refs.~\cite{Natsuume:2019sfp,Natsuume:2019vcv,Wu:2019esr,Balm:2019dxk,Ceplak:2019ymw,Liu:2020yaf,Ahn:2019rnq,Ahn:2020bks,Abbasi:2020ykq,Jansen:2020hfd,Ramirez:2020qer,Ahn:2020baf,Natsuume:2020snz,Kim:2020url,Sil:2020jhr,Ceplak:2021efc,Jeong:2021zhz,Natsuume:2021fhn,Blake:2021hjj,Kim:2021xdz,Wang:2022mcq,Amano:2022mlu,Yuan:2023tft,Grozdanov:2023txs,Natsuume:2023lzy,Jeong:2023zkf}).}
According to the AdS/CFT duality or the holographic duality \cite{Maldacena:1997re,Witten:1998qj,Witten:1998zw,Gubser:1998bc}, there are special points called ``pole-skipping points" in the complex momentum space $(\omega,q)$, where $\omega$ is the frequency and $q$ is the wave number.  At a pole-skipping point, the Green's function has the structure $G^R =0/0$, namely the residue of a pole vanishes so that the would-be pole is skipped. 

This is a universal property in the sense that the pole-skipping points $\omega$ are always located at Matsubara frequencies $\nw := \omega/(2\pi T) =(s-1)i $ ($s$ is the spin of the bulk field) and continue to $\nw_n = (s-1-n)i $
for a non-negative integer $n$. 

In the gravitational scalar mode (sound mode), they start from $\nw_{-1}=+i$. 
It is argued \cite{Grozdanov:2017ajz,Blake:2018leo} that the $\nw_{-1} = +i$ point is related to many-body quantum chaos \cite{Shenker:2013pqa,Roberts:2014isa,Roberts:2014ifa,Shenker:2014cwa,Maldacena:2015waa}. 

However, it remains unclear in general how the pole-skipping appears as a physical phenomenon.
This is because pole-skipping points are typically located in the complex momentum space $(\omega, q)$ which is outside the physical region. 
In this paper, we study the issue in the context of the AdS soliton \cite{Horowitz:1998ha} and show that 
\textit{the pole-skipping leaves its fingerprint in the normal mode spectrum.}

Namely, the normal mode appears as a pole in the Green's function, but the Green's function has the structure $G^R=0/0$ at a pole-skipping point so that some normal modes are actually skipped, \ie, they do not appear in the spectrum.  We call them ``missing states."


The AdS soliton is obtained by the double Wick rotation from the AdS black hole (\sect{preliminaries}). The AdS soliton is not a black hole. The geometry has a compact $S^1$-direction $z$ with periodicity $l$, and the geometry ends smoothly at the ``horizon." 

The pole-skipping is often discussed in a \bh background, but the pole-skipping occurs even in the AdS soliton background \cite{Natsuume:2023lzy}. Because of the double Wick rotation, there exist pole-skipping points at integer $\nq_z$ where $\nq_z:=q_z/(2\pi/l)$ and $q_z$ is the $S^1$-momentum.

The AdS soliton describes the confining phase of the dual gauge theory whereas the AdS \bh describes the plasma phase. 
Soon after the holographic duality was proposed, the normal modes in the AdS soliton background were computed \cite{Csaki:1998qr,deMelloKoch:1998vqw}. The results are compared with lattice computations of the pure Yang-Mills (YM) theory. The normal mode computation in  Refs.~\cite{Csaki:1998qr,deMelloKoch:1998vqw} uses the bulk massless scalar perturbation, but it was extended to the other bulk perturbations \cite{Brower:1999nj,Brower:2000rp}.%
\footnote{ One would say that the normal mode computation provides the first example of the real-world applications of the holographic duality. Since then, the duality has been applied to various areas such as QCD, condensed-matter physics, non-equilibrium physics, nonlinear physics, and quantum information (see, \eg, Refs.~\cite{CasalderreySolana:2011us,Natsuume:2014sfa,Ammon:2015wua,Zaanen:2015oix,Hartnoll:2016apf,Baggioli:2019rrs}). }

However, normal modes with $q_z\neq0$ are little discussed in the literature. 
When one computes normal modes in the AdS soliton background, one typically ignores all $S^1$ modes and $S^5$ modes. This is because one would like to compare with the spectrum of the pure $(2+1)$-dimensional YM theory. The $S^1$ mode and $S^5$ modes carry $U(1)_\text{KK}$ charge and the R-charges, respectively. The pure YM theory does not have such states. 

One hopes that they decouple in the $R\to0$ limit, where $l=2\pi R$.
However, the AdS soliton has the only 1 scale, the Kaluza-Klein scale $1/R$, so all states have the mass $O(1/R)$. 
In any case, the AdS \bh and the AdS soliton are dual to the $\mathcal{N}=4$ super-Yang-Mills (SYM) theory, not the pure YM theory. So, presumably the $\mathcal{N}=4$ SYM with $S^1$ should have the states from $S^1$ modes and $S^5$ modes. 

Recently, we derive the master equations in the AdS soliton background for various bulk perturbations such as the Maxwell perturbations and the gravitational perturbations \cite{Natsuume:2023lzy}. In this paper, we compute the spectrum of the $S^1$ modes. It turns out that the spectrum has some puzzling features (\sect{spectrum}). 
This is because the pole-skipping is not taken into account: the would-be states are missing at the pole-skipping points (\sect{pole-skip}).
The puzzling features are absent once one takes into account these pole-skipping points.

\section{Preliminaries}\label{sec:preliminaries}

\subsection{The AdS soliton}\label{sec:soliton}
The Schwarzschild-AdS$_5$ (SAdS$_5$) \bh is given by
\begin{subequations}
\label{eq:sads5}
\begin{align}
ds_5^2 &= r^2(-fdt^2+dx^2+dy^2+dz^2)+\frac{dr^2}{r^2f} \\
&= \frac{r_0^2}{u} (-fdt^2+dx^2+dy^2+dz^2)+\frac{du^2}{4u^2f}~, \\
f &= 1-\left(\frac{r_0}{r}\right)^4 = 1-u^2~, 
%
\end{align}
\end{subequations}
where $u:=r_0^2/r^2$.
For simplicity, we set the AdS radius $L=1$ and the horizon radius $r_0=1$. The Hawking temperature is given by $\pi T =r_0/L^2$. 

We compactify the $z$-direction as $0\leq z<l$. 
The AdS soliton is obtained by the double Wick rotation from the SAdS$_5$ black hole:
\begin{align}
t=i\hz~, \quad \hatt=iz~.
%
\end{align}
Then, the metric becomes
\begin{align}
ds_5^2 &= \frac{r_0^2}{u} (-d\hatt^2+dx^2+dy^2+fd\hz^2)+\frac{du^2}{4u^2f}~, 
\label{eq:ads_soliton}
\end{align}
with $f=1-u^2$. 

For the SAdS$_5$ black hole, the imaginary time direction has the periodicity $\beta=\pi/r_0$ to avoid a conical singularity. Similarly, for the AdS soliton, $\hz$ has the periodicity $l=\pi/r_0$. The AdS soliton is not a black hole. Rather, it has a cigar-like geometry, and the geometry ends smoothly at $u=1$ because of the factor $f$ just like the Euclidean black hole. We focus on the asymptotically AdS$_5$ geometry, but the generalization to the other dimensions is straightforward.

In order to distinguish the SAdS$_5$ and the AdS soliton, we use variables such as $\hatt$, but we omit ``~$\hat{~}$~" in the rest of our paper. 

\subsection{Boundary condition}

Let us consider perturbations in the background. For the SAdS$_5$, there is a $SO(3)$ invariance for the boundary direction $(x,y,z)$, so one can set the perturbation of the form 
\begin{align}
%
Z(u) e^{-i\omega t+iqx}
\end{align}
without loss of generality. 
The field equation typically takes the form
\begin{align}
0 \sim Z''+\frac{1}{u-1}Z'+\frac{\nw^2}{4(u-1)^2}Z~, \quad(u\to1)~,
%
\end{align}
where $'=\del_u$ and $\nw:=\omega/(2\pi T)=\omega/2$. One imposes the incoming-wave boundary condition near the horizon, namely 
\begin{align}
Z \propto (u-1)^{-i\nw/2}~.
\label{eq:incoming}
\end{align}
Because a perturbation is absorbed by the black hole, one obtains quasinormal modes, namely poles are located in the complex $\omega$-plane. 

For the AdS soliton, the $SO(3)$ invariance is broken due to the $S^1$-direction $z$, so we consider the perturbation of the form
\begin{align}
Z(u) e^{-i\omega t+iq_x x+iq_z z} = Z(u)e^{ip_ix^i+iq_z z}~,
%
\end{align}
where $p_i=(-\omega,q_x,0)$
However, there is a remaining $SO(1,2)$ invariance for $(t,x,y)$, so $\omega$ and $q_x$ appears only in the combination $\ki^2:=-\omega^2+q_x^2$. 
Also, the $z$-direction is compact, so $q_z$ takes only a discrete value:
\begin{align}
q_z = \frac{2\pi n}{l} \text{~~or~~} \nq_z:=\frac{q_z}{\frac{2\pi}{l}} =n
%
\end{align}
for an integer $n$. 

The AdS soliton does not have a horizon, and the geometry smoothly ends at $u=1$, so one obtains normal modes instead of quasinormal modes. The field equation typically takes the form
\begin{align}
0 \sim Z''+ \frac{1}{u-1}Z'-\frac{\nq_z^2}{4(u-1)^2}Z~,
\quad(u\to1)~,
\label{eq:near_horizon}
\end{align}
near the tip of cigar, where $\mfq_z:=q_z/(2\pi/l)=q_z/2$, so there are 2 solutions:
\begin{align}
Z \propto (u-1)^\lambda~, \quad \lambda=\pm \nq_z/2~.
%
\end{align}
In Ref.~\cite{Natsuume:2023lzy}, we propose the boundary condition as
\begin{align}
Z \propto (u-1)^{\nq_z/2}~, \quad(\nq_z>0)~.
\label{eq:bc_soliton}
\end{align}
Note that the above choice is the analytic continuation from the \bh case \eqref{eq:incoming}, but this is not the reason why we choose the boundary condition. 
We impose the boundary condition that the perturbations are regular at the tip of cigar $u=1$. The mode $(u-1)^{\nq_z/2}$ is regular, but the mode $(u-1)^{-\nq_z/2}$ diverges at the tip. Another way to justify the boundary condition comes from quantum mechanics. The field equation can be rewritten as a Schr\"{o}dinger problem with angular momentum, and this is the choice in the standard textbook treatment of quantum mechanics (Chap.~35 of \cite{LL}). 

Because $\nq_z$ is the $S^1$-momentum, $\nq_z<0$ is also possible. In this case, one chooses $Z \propto (u-1)^{-\nq_z/2}$. We set $\nq_z>0$ for simplicity.

Now, consider the boundary condition at $u=0$. The asymptotic behavior depends on the field one considers. For example, consider the massless scalar field $0=\nabla^2\phi$. The field has the asymptotic behavior 
\begin{align}
\phi \sim A+Bu^2~, \quad (u\to0)~.
%
\end{align}
According to the standard AdS/CFT dictionary, the Green's function is given by
\begin{align}
G^R \propto \frac{B}{A}~.
\label{eq:pole}
\end{align}
So, a pole corresponds to $A=0$ (if $B\neq0$). Then, if one is interested in the spectrum, it is enough to solve the perturbation equation under the boundary condition $A=0$. We take this boundary condition and compute the poles of the Green's function instead of the entire Green's function.

\subsection{Tensor decomposition}

In this paper, we consider the scalar field, the Maxwell field,%
\footnote{One can regard the Maxwell field as the one which couples to the $U(1)$ subgroup of the $SU(4)_R$ R-symmetry of the $\mathcal{N}=4$ SYM. In general, the R-symmetry is anomalous, but one can choose a $U(1)$ subgroup which is not anomalous.}
and gravitational perturbations. In the SAdS$_5$ background, one decomposes perturbations under the $SO(2)$ transformation $(y,z)$. The scalar (vector) mode transforms as a scalar (vector) under the transformation. For example, the Maxwell perturbations $A_M$ are decomposed as
\begin{subequations}
\begin{align}
\text{scalar mode (diffusive mode): }& A_t~, A_u~, A_x~, \\
\text{vector mode: }& A_y~, A_z~.
%
\end{align}
\end{subequations}
One would fix the gauge $A_u=0$, but we do not fix the gauge and carry out analysis in a fully gauge-invariant manner 
[both under the $U(1)$ gauge transformation and under the diffeomorphism]. 
This is the formalism developed by Kodama and Ishibashi \cite{Kodama:2003jz}. Similarly, gravitational perturbations are decomposed as the scalar mode (sound mode), the vector mode (shear mode), and the tensor mode.

The vector mode $A_y, A_z$ is gauge invariant by themselves. The scalar mode has 3 perturbations, but one is redundant due to the gauge symmetry. One can define 2 gauge-invariant variables $\mfA_a$ where $x^a=(t,u)$.

Similarly, in the AdS soliton background, we decompose perturbations under the $SO(1,2)$ transformation $y^i=(t,x,y)$. The scalar (vector) mode transforms as a scalar (vector) under the transformation. 
For example, the Maxwell perturbations $A_M$ are decomposed as
\begin{subequations}
\begin{align}
\text{scalar mode: }& A_z~, A_u~, A_L~, \\
\text{vector mode: }& A_y~,
%
\end{align}
\end{subequations}
where $A_L$ is the longitudinal mode.

\section{The spectrum}\label{sec:spectrum}

\subsection{Master equations}\label{sec:master_eq}

In Ref.~\cite{Natsuume:2023lzy}, we derive the master equations in the AdS soliton background for various bulk perturbations. It turns out that the master equations coincide with the master equations for the SAdS$_5$ black hole  \cite{Kovtun:2005ev} after the double Wick rotation. Below we summarize the master equations for reader's convenience.

\subsubsection{Massless scalar field (gravitational tensor)}
The field equation is given by
\begin{align}
0 =\phi''+\left(\frac{f'}{f} - \frac{1}{u}\right)\phi' - \frac{ \nq_z^2+\nk^2 f }{uf^2} \phi~,
\label{eq:scalar}
\end{align}
where $\nk:=p/(2\pi/l)=p/2$. Asymptotically, $\phi \sim A+B u^2$. Just like the \bh case, one can show that the gravitational tensor mode takes the form of the massless scalar field.

\subsubsection{Maxwell vector}
The field equation is given by
\begin{align}
0 &= A_y''+ \frac{f'}{f}A_y' - \frac{ \nq_z^2+\nk^2f }{uf^2} A_y~.
%
\end{align}
Asymptotically, $A_y \sim A+B u$.

\subsubsection{Maxwell scalar}
The gauge-invariant variables are given by
\begin{subequations}
\begin{align}
\mfA_z &= A_z - iq_z A_L~, 
\\ 
\mfA_u &= A_u - A_L'~.
%
\end{align}
\end{subequations}
The Maxwell equation becomes
\begin{subequations}
\begin{align}
0 &= \frac{i\nq_z}{2uf}\mfA_z + (f\mfA_u)'~, 
\label{eq:eom_Au} \\
0 &= (2\nq_z^2+\nk^2f) \mfA_u + i\nq_z\mfA_z'~.
%
\end{align}
\end{subequations}
We choose $\mfA_z$ as the master variable.
Then, the master equation is given by
\begin{align}
0 &= \mfA_z''+\frac{\nq_z^2f'}{(\nq_z^2+\nk^2f)f}\mfA_z' - \frac{ \nq_z^2+\nk^2f }{uf^2} \mfA_z~.
\label{eq:master_diffusive}
\end{align}
Asymptotically, $\mfA_z \sim A+B u$.

\subsubsection{Gravitational vector}
The gauge-invariant variables are given by
\begin{subequations}
\begin{align}
\mfh_{zy} &= h_{zy}^{(1)} - iq_z h_y^{(1)}~, 
\\ 
\mfh_{uy} &= h_{uy}^{(1)} - \frac{1}{u} (u h_y^{(1)})'~.
%
\end{align}
\end{subequations}
The Einstein equation becomes
\begin{subequations}
\begin{align}
0 &= \frac{i\nq_z}{2uf}\mfh_{zy}+(f\mfh_{uy})'~, 
\label{eq:eom_huy} \\
0 &= -\frac{2iu}{\nq_z} (\nq_z^2+\nk^2f) \mfh_{uy}+(u\mfh_{zy})'~.
%
\end{align}
\end{subequations}
We choose $\mfh_{zy}$ as the master variable.
Then, the master equation is given by
\begin{align}
0 &= \Zone''
-\frac{(\nq_z^2+\nk^2f)f-\nq_z^2uf'}{uf(\nq_z^2+\nk^2f)} \Zone' 
-\frac{\nq_z^2+\nk^2f}{uf^2}\Zone~,
\label{eq:master_shear}
\end{align}
where $\Zone=u\mfh_{zy}$. 
Asymptotically, $\Zone \sim A+B u^2$.

\subsubsection{Gravitational scalar}
The field equation is given by
\begin{align}
0 &= \Zone'' 
- \frac{-3\nq_z^2(1+u^2)+\nk^2(-3+2u^2-3u^4)}{uf \{-3\nq_z^2+\nk^2(-3+u^2)\} }\Zone' 
\nonumber \\
&+ \frac{3\nq_z^4+\nk^4(3-4u^2+u^4)+\nk^2\{ \nq_z^2(6-4u^2)-4u^3f \} }{ uf^2\{-3\nq_z^2+\nk^2(-3+u^2)\} } \Zone~,
\label{eq:master_sound}
\end{align}
where $\Zone= u \{ \mfh_{zz} - (f-uf') \mfh_L \}$. Asymptotically, $\Zone \sim A+B u^2$.

\subsection{Methods}

All master variables behave as \eq{near_horizon}:
\begin{align}
0 \sim \Zone''+ \frac{1}{u-1}\Zone'-\frac{\nq_z^2}{4(u-1)^2}\Zone~,
\quad(u\to1)~,
%
\end{align}
near the tip of cigar. Thus, we set the ansatz
\begin{align}
\Zone=(1-u^2)^{\nq_z/2}\Za~.
%
\end{align}
To solve the equation for $\Za$, there are several methods. Among them, the popular ones are
\begin{enumerate}
\item Frobenius method
\item Shooting method
\end{enumerate}
We mainly use Frobenius method but use both in this paper. 

In Frobenius method, the solution obeying the boundary condition at $u=1$ is obtained by the power series expansion around $u=1$:
\begin{align}
\Za = \sum_{n=0}^\infty a_n (u-1)^n~.
%
\end{align}
We are interested in normal modes, so we impose the source-free condition or no slow falloff condition at $u=0$ as described in \eq{pole}:
\begin{align}
\Za|_{u=0} = \sum_{n=0}^N (-)^n a_n =0~.
%
\end{align}
One truncates the series after a large number of terms $n=N$. One can check the accuracy as one goes to higher series. 
 
This Frobenius method is actually a little naive because some master equations have regular singular points 
in the region $0<u<1$ (for the Maxwell scalar, gravitational vector, and gravitational scalar modes). 
This means that the convergence of the power-series solutions is not guaranteed at $u=0$ (\appen{alternative}). In order to check our results, 
\begin{enumerate}
\item
We use Frobenius method for alternative master variables where this problem does not occur (for Maxwell scalar and gravitational vector modes).
\item
We also use the shooting method for all modes. 
\end{enumerate}
The results obtained by these methods agree with the naive Frobenius method. 

\subsection{Results}\label{sec:results}

\begin{table}[tb]
\begin{center}
\begin{tabular}{l|r|cccc}
&& $n=0$ & $n=1$ & $n=2$ & $n=3$ \\
\hline
Maxwell vector	& $\nq_z=0$ & 5.131 & 22.48 & 51.21 & 91.41 \\
				& $=1$ & 14.44 & 37.43 & 71.88 & 117.8 \\
				& $=2$ & 30.89 & 59.41 & 99.52 & 151.2 \\
				& $=3$ & 54.83 & 88.63 & 134.3 & 191.5 \\
Maxwell scalar	& $\nq_z=0$ & 11.59 & 34.53 & 68.97 & 114.9 \\
				&			& $\vee$& $\vee$&$\vee$&$\vee$ \\
				& $=1$ & 7.162 & 23.80 & 52.40 & 92.55 \\
				& $=2$ & 21.45 & 43.22 & 77.33 & 123.1 \\
				& $=3$ & 43.36 & 69.96 & 109.4 &160.8 \\
\hline
Gravitational tensor& $\nq_z=0$ & 11.59 & 34.53 & 68.97 & 114.9 \\
(massless scalar)	& $=1$ & 23.47 & 52.25 & 92.47 & 144.2 \\
				& $=2$ & 42.36 & 76.91 & 122.9 & 180.3 \\
				& $=3$ & 68.60 & 108.7 & 160.3 & 223.4 \\
Gravitational vector			& $\nq_z=0$ & 18.68 & 47.50 & 87.72 & 139.4 \\
				&			& $\vee$& $\vee$&$\vee$&$\vee$ \\
				& $=1$ & 10.79 & 33.87 & 68.29 & 114.2 \\
				& $=2$ & 28.04 & 56.20 & 96.14 & 147.7 \\
				& $=3$ & 52.47 & 85.74 & 131.1 & 188.2 \\
Gravitational scalar			& $\nq_z=0$ & 5.457 & 30.44 & 65.12 & 111.1 \\
				& $=1$ & 20.13 & 48.68 & 88.85 & 140.5 \\
				&			& $\wr$& $\vee$&$\vee$&$\vee$ \\
				& $=2$ & 20.27 & 40.32 & 74.00 & 119.6 \\
				& $=3$ & 42.27 & 67.49 & 106.4 & 157.5 
\end{tabular}
\caption{$m_3^2$ for the first 4 states obtained by numerical computations.}
\label{table:spectrum1}
\end{center}
\end{table}

In \tabl{spectrum1}, we show the results by Frobenius method. 
Here, $m_3$ is the mass spectrum in the dual $(2+1)$-dimensional gauge theory, \ie, $m_3^2:=-p^2=-4\nk^2$.
\begin{enumerate}
\item
First, note that the tensor mode (massless scalar field) and the Maxwell scalar mode have the same spectrum when $\nq_z=0$. In fact, field equations are identical under an appropriate transformation (\sect{discussion}).
\item
In addition, the spectrum has some puzzling features:
\begin{enumerate}
\item
Let us consider a fixed $n$. One would expect that $m_3$ increases as one increases $\nq_z$:
\begin{align}
m_3(\nq_z=0)<m_3(\nq_z=1)
\nonumber \\
<m_3(\nq_z=2)<\cdots~.
%
\end{align}
Namely, a Kaluza-Klein state with a larger $S^1$ momentum has a larger mass.
This is true for the Maxwell vector mode and the gravitational tensor mode, but it is not always the case. For the Maxwell scalar mode, $m_3(\nq_z=1)<m_3(\nq_z=0)$ for all $n$. A similar remark applies to the gravitational vector and scalar modes. 

\item
Let us compare different spins. One would expect that $m_3$ increases for a larger spin, namely for gravitational perturbations, 
\begin{align}
m_S<m_V<m_T~,
%
\end{align}
where $m_S,m_V,m_T$ are the mass of spin 0, 1, 2 states, respectively. This is true for $\nq_z=2,3$, but it is not always the case: $m_S<m_T<m_V$ for $\nq_z=0$,%
\footnote{This was pointed out previously in Ref.~\cite{Brower:2000rp}.}
and $m_V<m_S<m_T$ for $\nq_z=1$. For the Maxwell perturbations, $m_V<m_S$ for $\nq_z=0$.
\end{enumerate}
\end{enumerate}
These puzzling features suggest that there may be some problem in the structure of the table. 
In next section, we argue that the puzzling features come from the fact that we have not taken the pole-skipping into account.

\begin{table}[tb]
\begin{center}
\begin{tabular}{l|r|cccc}
&& $n=0$ & $n=1$ & $n=2$ & $n=3$ \\
\hline
Maxwell vector	& $\nq_z=0$ & 5.131 & 22.48 & 51.21 & 91.41 \\
				& $=1$ & 14.44 & 37.43 & 71.88 & 117.8 \\
				& $=2$ & 30.89 & 59.41 & 99.52 & 151.2 \\
				& $=3$ & 54.83 & 88.63 & 134.3 & 191.5 \\
Maxwell scalar	& $\nq_z=0$ & (0) & 11.59 & 34.53 & 68.97 \\
				& $=1$ & 7.162 & 23.80 & 52.40 & 92.55 \\
				& $=2$ & 21.45 & 43.22 & 77.33 & 123.1 \\
				& $=3$ & 43.36 & 69.96 & 109.4 &160.8 \\
\hline
Gravitational tensor& $\nq_z=0$ & 11.59 & 34.53 & 68.97 & 114.9 \\
				& $=1$ & 23.47 & 52.25 & 92.47 & 144.2 \\
				& $=2$ & 42.36 & 76.91 & 122.9 & 180.3 \\
				& $=3$ & 68.60 & 108.7 & 160.3 & 223.4 \\
Gravitational vector& $\nq_z=0$ & (0) & 18.68 & 47.50 & 87.72 \\
				& $=1$ & 10.79 & 33.87 & 68.29 & 114.2 \\
				& $=2$ & 28.04 & 56.20 & 96.14 & 147.7 \\
				& $=3$ & 52.47 & 85.74 & 131.1 & 188.2 \\
Gravitational scalar& $\nq_z=0$ & (0) & 5.457 & 30.44 & 65.12 \\
				& $=1$ & (6.000) &20.13 & 48.68 & 88.85 \\ 
				& $=2$ & 20.27 & 40.32 & 74.00 & 119.6 \\
				& $=3$ & 42.27 & 67.49 & 106.4 & 157.5 
\end{tabular}
\caption{The modified spectrum augmented by pole-skip points [the states in parentheses ``(~)"]. The pole-skip points are not part of the spectrum.}
\label{table:spectrum_mod}
\end{center}
\end{table}

\begin{table}[tb]
\begin{center}
\begin{tabular}{r|cccc}
&$n=0$ & $n=1$ & $n=2$ & $n=3$ \\
\hline
Maxwell vector	
& 5.131 & 22.48 & 51.21 & 91.41 \\
 scalar	
& (0) & 11.59 & 34.53 & 68.97 \\
\hline
Gravitational tensor			
& 11.59 & 34.53 & 68.97 & 114.9 \\
 vector			
& (0) & 18.68 & 47.50 & 87.72 \\
 scalar
& (0) & 5.457 & 30.44 & 65.12 
\end{tabular}
\caption{The $\nq_z=0$ spectrum. For gravitational perturbations, $m_S<m_V<m_T$ if one takes the pole-skip into account, but $m_S<m_T<m_V$ without pole-skip. For Maxwell perturbations, $m_S<m_V$ with pole-skip, but $m_V<m_S$ without pole-skip.}
\label{table:qz0}
\vskip2mm
\begin{tabular}{r|cccc}
&$n=0$ & $n=1$ & $n=2$ & $n=3$ \\
\hline
Maxwell vector	
& 14.44 & 37.43 & 71.88 & 117.8 \\
 scalar	
& 7.162 & 23.80 & 52.40 & 92.55 \\
\hline
Gravitational tensor			
& 23.47 & 52.25 & 92.47 & 144.2 \\
 vector			
& 10.79 & 33.87 & 68.29 & 114.2 \\
 scalar
& (6.000) &20.13 & 48.68 & 88.85  
\end{tabular}
\caption{The $\nq_z=1$ spectrum. For gravitational perturbations, $m_S<m_V<m_T$ if one takes the pole-skip into account, but $m_V<m_S<m_T$ without pole-skip. For Maxwell perturbations, $m_S<m_V$.}
\label{table:qz1}
\vskip2mm
\begin{tabular}{r|cccc}
&$n=0$ & $n=1$ & $n=2$ & $n=3$ \\
\hline
Maxwell vector	
& 30.89 & 59.41 & 99.52 & 151.2 \\
 scalar	
& 21.45 & 43.22 & 77.33 & 123.1 \\
\hline
Gravitational tensor			
& 42.36 & 76.91 & 122.9 & 180.3 \\
 vector			
& 28.04 & 56.20 & 96.14 & 147.7 \\
 scalar
& 20.27 & 40.32 & 74.00 & 119.6 			
\end{tabular}
\caption{The $\nq_z=2$ spectrum.  $m_S<m_V<m_T$. }
\label{table:qz2}
\end{center}
\end{table}

\section{Pole-skipping}\label{sec:pole-skip}

The pole-skipping is a new universal property of Green's functions. 
In the \bh case, there are special points called ``pole-skipping points" in the complex momentum space $(\omega,q)$. Near a pole-skipping point, a Green's function typically takes the form 
\begin{align}
G^R \propto \frac{\delta\omega+\delta q}{\delta\omega-\delta q}~.
\label{eq:Green_typical}
\end{align}
In this sense, the Green's function is not uniquely determined, and it depends on the slope $\delta q/\delta \omega$ how one approaches the point. 

The pole-skipping points $\omega$ are always located at Matsubara frequencies $\nw := \omega/(2\pi T) =(s-1)i $ ($s$ is the spin of the bulk field) and continue to $\nw_n = (s-1-n)i $
for a non-negative integer $n$. 
In the gravitational scalar mode (sound mode), they start from $\nw_{-1}=+i$. 
It is argued that the $\nw_{-1} = +i$ point is related to many-body quantum chaos. 

However, it remains unclear in general how the pole-skipping appears as a physical phenomenon.
This is because pole-skipping points are typically located in the complex momentum space $(\omega, q)$ which is outside the physical region. 

One would take the physical region as $\omega$ in the lower-half plane and real $q$. 
For the pole-skipping points $\nw=-i$,
\begin{enumerate}
\item
 The Maxwell scalar and the gravitational vector modes lie in the physical region (real $q$).
\item
The Maxwell vector, the gravitational tensor, and gravitational scalar modes do not lie in the physical region.
\end{enumerate}
Note that the above physical region would exclude the chaotic pole-skipping at $\nw=+i$, so it is subtle to exclude pole-skipping points outside the physical region. 

Now, the pole-skipping is often discussed in a \bh background, but the pole-skipping occurs even in the AdS soliton background \cite{Natsuume:2023lzy}. Because of the double Wick rotation, 
\begin{enumerate}
\item
The universality of the pole-skipping points $\omega$ is translated to the universality of the pole-skipping points $\nq_z$.  
\item
The pole-skipping points start from $\nq_z=s-1$ and continue to $\nq_z=s-1-n$ $(n=1,2,\cdots)$. 
\end{enumerate}
Note that $\nq_z$ takes only discrete values physically $\nq_z\in \mathbb{Z}$, which coincide with the pole-skipping points. This is reminiscent of the fact that pole-skipping points in black holes are located at Matsubara frequencies.

As a simple example, consider the massless scalar field $\phi$ and find the first pole-skipping point $\nq_z=1$. 
%
%
The solution can be written as a power series:
\begin{align}
\Za(u) = \sum_{n=0}\, a_n\, (u-1)^{n}~.
%
\end{align}
At the lowest order, one obtains
\begin{align}
0=\frac{1}{4}(2\nk^2+3\nq_z^2)a_0+(1+\nq_z)a_1~.
%
\end{align}
Normally, this equation determines $a_1$ from $a_0$. However, when $(\nq_z,\nk^2)=(-1,-3/2)$, both $a_0$ and $a_1$ are free parameters, and the bulk solution is not uniquely determined. As a result, the dual Green's function is not uniquely determined. See Ref.~\cite{Natsuume:2023lzy} for systematic analysis.

It turns out that most pole-skipping points lie outside the physical region like the \bh case. We take $\nq_z>0$, but most pole-skipping points are located at $\nq_z=-n<0$.

However, some pole-skipping points lie inside the physical region. For the Maxwell scalar, the gravitational vector, and the gravitational scalar modes, there is a pole-skipping point at
\begin{align}
(\nq_z,m_3^2)=(0,0)~,
%
\end{align}
which corresponds to the ``hydrodynamic" pole-skipping in the \bh case. The gravitational scalar mode  has another pole-skipping point at 
\begin{align}
(\nq_z,m_3^2)=(1,6)~,
%
\end{align}
which corresponds to the ``chaotic" pole-skipping in the \bh case.

The spectrum in \tabl{spectrum1} looks more ``natural" if one takes pole-skipping points into account (\tabl{spectrum_mod}). The states in parentheses ``(~)" are pole-skipping points, and we call them ``missing sates."
In this table,
\begin{enumerate}
\item
For a fixed $n$, $m_3$ always increases as one increases $\nq_z$.
\item
$m_3$ always increases for a larger spin (Tables~\ref{table:qz0}, \ref{table:qz1}, and \ref{table:qz2}).
\end{enumerate}

\begin{figure}[tb]
\centering
\scalebox{0.38}{ \includegraphics{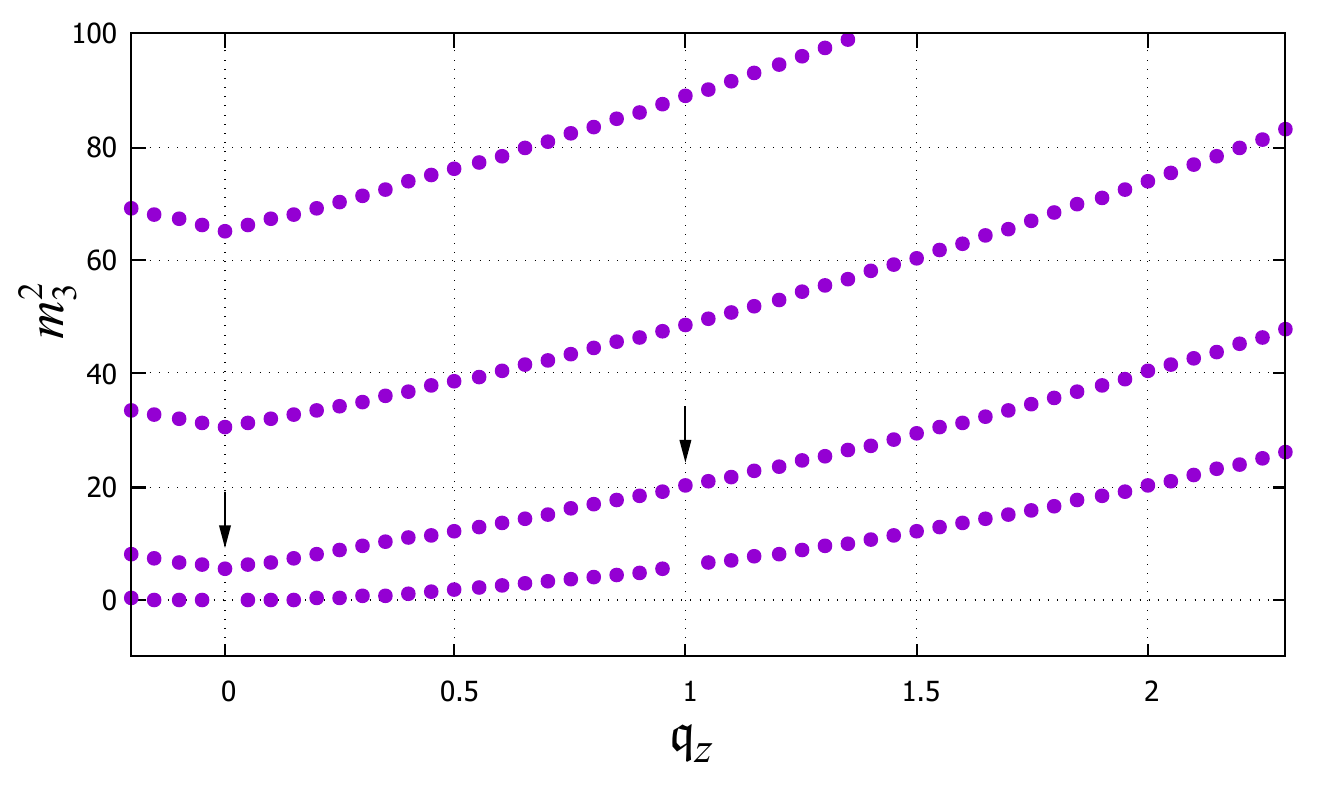} } 
\vskip2mm
\caption{The normal modes of the gravitational scalar mode for continuous $\nq_z$. One can see the ``missing states" at ``hydrodynamic" and ``chaotic" pole-skip points.
}
\label{fig:spectrum}
\end{figure}

The ``missing states" can be most easily seen by making $\nq_z$ continuous and approaching pole-skipping points. \fig{spectrum} shows the result for the gravitational scalar mode:
\begin{enumerate}
\item
One can indeed see that states are missing at the hydrodynamic pole-skipping $(\nq_z,m_3^2)=(0,0)$ and at the chaotic pole-skipping $(\nq_z,m_3^2)=(1,6)$.
\item
The $(\nq_z,m_3^2)\approx(0,5.457)$ state was part of the $n=0$ spectrum in \tabl{spectrum1}, but it merges with the $n=1$ state continuously (the arrow in \fig{spectrum}). 
This suggests that the state is actually the part of the $n=1$ spectrum. A similar remark applies to the $(\nq_z,m_3^2)\approx(1,20.13)$ state. 
\end{enumerate}

In the \bh case, 
the Green's function \eqref{eq:Green_typical}
%
%
depends on the slope $\delta q/\delta \omega$ how one approaches the pole-skipping point. In the AdS soliton case, the situation is different however. In this case, $\nq_z$ is actually discrete, so one first fixes $\nq_z=0$, and so on, so one cannot choose the slope. Instead, the pole-skipping appears as ``missing states"%
\footnote{See Ref.~\cite{Natsuume:2019xcy} for the early interpretation along this line.}
: the would-be pole is skipped.

%
%

\section{At pole-skipping points}

We argued that the pole-skipping appears as ``missing states."
A natural question is whether they actually exist as normal modes or not. In order to answer to the question, we take the following 2 issues into account and solve the field equations directly at pole-skipping points:
\begin{enumerate}
\item
The first issue is the master equations. The pole-skipping limits of master equations are subtle, so we go back to the Maxwell equation and the Einstein equation and derive the field equations directly at pole-skipping points. 
\item
The second issue is the boundary condition at the tip. We impose the boundary condition \eqref{eq:bc_soliton}, but at pole-skipping points, the near-horizon behavior of master equations change, so one has to reexamine the issue. Also, the boundary condition \eqref{eq:bc_soliton} assumes $\nq_z\neq0$, so we need to specify the boundary condition when $\nq_z=0$. 
\end{enumerate}
When $\nq_z=0$, one usually imposes the Neumann boundary condition $Z'|_{u=1}=0$. But it turns out whether there exist normal modes or not at pole-skipping points depends crucially on the boundary condition, so we discuss it a little carefully.

We are interested in normal modes or source-free solutions. We show that there is no nontrivial normal mode at pole-skipping points under our boundary condition. 

\subsection{Revisiting the boundary condition at the tip}

Near the tip of the cigar $u=1$, we take the coordinate system which is regular at $u=1$ and impose the boundary condition that perturbations are regular in the coordinate system. 

Consider the following metric:
\begin{align}
ds^2 = \frac{dr^2}{F(r)}+F(r)dz^2+\cdots.
%
\end{align}
Suppose $F(r) = F'(1)(r-1)+\cdots$ near the tip $r=1$. As is well-known, the metric takes the form of polar coordinates:
\begin{subequations}
\begin{align}
ds^2 &\sim d\rho^2+\rho^2d\theta^2~, \\
\rho &= \int \frac{dr}{\sqrt{F}} \sim 2\left(\frac{r-1}{F'}\right)^{1/2}~, \\
\theta &= \frac{F'}{2} z~.
%
\end{align}
\end{subequations}
There is no conical singularity at $\rho=0$ when $\theta$ has the periodicity $2\pi$ or $z$ has the periodicity $l=4\pi/F'(1)$. For the AdS soliton, $F(r)=r^2(1-r^{-4})$, so $l=\pi$.

The point $\rho=0$ is a coordinate singularity, so we further make the coordinate transformation:
\begin{align}
X=\rho \cos\theta~, Y=\rho\sin\theta~.
%
\end{align}
The coordinates $(X,Y)$ are regular coordinates. 

For example, consider the Maxwell field $A_M$. Under the coordinate transformation,
\begin{subequations}
\begin{align}
A_z &= \frac{F'}{2}A_\theta \sim \sqrt{F} (-A_X\sin\theta + A_Y\cos\theta)~,\\
A_r &= \frac{\del\rho}{\del r}A_\rho \sim \frac{1}{\sqrt{F}} (A_X\cos\theta + A_Y\sin\theta)~.
%
\end{align}
\end{subequations}
Because the coordinates $(X,Y)$ are regular at $r=1$, $(A_X,A_Y)$ must be regular there. This implies 
\begin{align}
A_z \sim O(F^{1/2})~, \quad A_r \sim O(F^{-1/2})~.
%
\end{align}
It is easy to extend the analysis to our metric and extend to the other perturbations. The boundary condition is
\begin{center}
A perturbation with a lower index $z$ $(u)$ \\
has the factor $f^{1/2}$ ($f^{-1/2}$). 
\end{center}
or
\begin{align}
\begin{array}{lll}
\text{Maxwell scalar: } &\mfA_z \sim O(f^{1/2})~, &\mfA_u \sim O(f^{-1/2})~, \\
\text{Gravitational vector: }  &\mfh_{zi} \sim O(f^{1/2})~, &\mfh_{ui} \sim  O(f^{-1/2})~, \\
\text{Gravitational scalar: } &\mfh_{zz} \sim O(f)~, &\mfh_{zu} \sim O(1)~, \\
&\mfh_{uu} \sim O(f^{-1})~. &
\end{array}
\label{eq:bc1}
\end{align}

\subsection{``Hydrodynamic" pole-skip}

The master equations in \sect{master_eq} become singular at $(\nq_z,\nk^2)=(0,0)$, so we go back to the Maxwell equation and the Einstein equation and derive field equations directly at pole-skipping points.

\begin{enumerate}
\item
Maxwell scalar:
When $\nq_z=\nk^2=0$,  $\mfA_z$ and $\mfA_u$ decouple, and the $\mfA_z$ equation becomes
\begin{align}
0&=\Zone''~, \quad \Zone=\mfA_z~. 
%
\end{align}
The $\Zone$ equation coincides with the original master equation by taking the $\nq_z=0$ limit first and then taking the $\nk^2=0$ limit (not the other way around). The solution is
\begin{align}
\Zone &=A+Bu~. 
%
\end{align}
Imposing our boundary condition at the tip, one gets $\Zone=A(1-u)$, but we also impose the source-free condition $A=0$ asymptotically, so there is no solution.
The $\mfA_u$ equation is $0= (f\mfA_u)'$ whose solution is $\mfA_u\propto (1-u^2)^{-1}$, 
so the solution does not satisfy the boundary condition.

\item
Gravitational vector:
When $\nq_z=\nk^2=0$,  $\mfh_{zy}$ and $\mfh_{uy}$ decouple, and the $\mfh_{zy}$ equation becomes
\begin{align}
0&=\Zone''-\frac{\Zone'}{u}~, \quad \Zone= u\mfh_{zy}~. 
%
\end{align}
The $\Zone$ equation coincides with the original master equation by taking the $\nq_z=0$ limit first and then taking the $\nk^2=0$ limit. The solution is 
\begin{align}
\Zone &=A+Bu^2~.
%
\end{align}
Imposing our boundary condition at the tip, one gets $\Zone=A(1-u^2)$, but we also impose the source-free condition $A=0$ asymptotically, so there is no solution.
The $\mfh_{uy}$ equation is $0= (f\mfh_{uy})'$ whose solution is $\mfh_{uy}\propto (1-u^2)^{-1}$, 
so the solution does not satisfy the boundary condition.

\item
Gravitational scalar:
When $\nq_z=\nk^2=0$,  $\mfh_{zu}$ decouples. 
The master equation is given by
\begin{subequations}
\begin{align}
0 &=\Zone'-\frac{2u}{1+u^2}\Zone \\
\to \Zone &= A(1+u^2)~,
%
\end{align}
\end{subequations}
where $\Zone$ is the original master variable. Imposing the source-free condition gives $A=0$, so there is no solution. 

On the other hand, if one takes the $\nq_z=0$ limit first and then takes the $\nk^2=0$ limit in the original master equation, one obtains
\begin{align}
0=\Zone''+ \frac{3-2u^2+3u^4}{uf(-3+u^2)}\Zone' 
\nonumber \\
- \frac{4u^2}{f(-3+u^2)}\Zone~.
\label{eq:master_sound0}
\end{align}
These 2 equations are related to each other. Setting 
\begin{subequations}
\begin{align}
F &= \Zone'-\frac{2u}{1+u^2}\Zone~,\\
G &=-\frac{3-5u^2+9u^4+u^6}{uf(1+u^2)(3-u^2)}~,
%
\end{align}
\end{subequations}
the equation
\begin{align}
0 &= F'+GF
%
\end{align}
reduces to the original master equation \eqref{eq:master_sound0}.

Solving the other components gives
\begin{subequations}
\begin{align}
\mfh_{zu} &= \frac{C}{1-u^2}~, \\
\mfh_{L} &= \frac{C_1}{u}+C_2~, \\
\mfh_{zz} &= (1+u^2)\left(\frac{C_1}{u}+C_2\right)~, \\
\mfh_{uu} &= \frac{C_1u-C_2}{2u(1-u^2)}~.
%
\end{align}
\end{subequations}
$\mfh_{zu}$ has no regular solution, and $C=0$. From the boundary condition at $u=1$, $\mfh_{zz}\sim O(u-1)$, so $C_2=-C_1$. But we also impose the source-free condition asymptotically, so $C_1=0$, and there is no solution.

\end{enumerate}

\subsection{``Chaotic" pole-skip}

The master equation coincides with the original master equation:
\begin{align}
0=\Zone''-\frac{1-3u^2}{uf}\Zone'+ \frac{1-3u^2+8u^3}{2uf^2}\Zone~. 
%
\end{align}
Near the tip $u\to1$,
\begin{align}
\Zone=(1-u^2)^\lambda \tilZ \quad \to \lambda=1/2,3/2~.
%
\end{align}
Note $\lambda=\pm1/2$ for a generic $(\nq_z,\nk^2)$, and we choose $\lambda=1/2$. But at the pole-skipping point, the near-horizon behavior of the master equation changes so that we obtain a different indicial equation. This is well-known for the chaotic pole-skipping for black holes. 


Choosing $\lambda=1/2$, one gets
\begin{align}
0= \tilZ''-\frac{1}{u}\tilZ'+\frac{1+2u}{2u(1+u)^2}\tilZ~.
%
\end{align}
Setting $x=1+u$ and 
\begin{align}
\tilZ = x^{(1+i)/2}\mathcal{Z}(x)~,
%
\end{align}
one obtains the hypergeometric differential equation:
\begin{subequations}
\begin{align}
0 &= x(1-x)\del_x^2 \mathcal{Z} + \{ c-(1+a+b)x\} \del_x \mathcal{Z} - ab \mathcal{Z}~, \\
a &=b= \frac{-1+i}{2}~, \quad c=1+i~.
%
\end{align}
\end{subequations}
Then, the solution is given by the hypergeometric functions:
\begin{align}
\tilZ = c_1 x^{(1+i)/2}{}_2F_1(a,a,c;x)+ c_2x^{(1-i)/2}{}_2F_1(a^*,a^*,c^*;x)~.
%
\end{align}

Asymptotically, $\tilZ\sim A+Bu^2$, where $A,B$ are some linear combination of $c_1,c_2$. From the source-free condition, $c_2$ is written by $c_1$:
\begin{align}
c_2 = -c_1 \frac{\Gamma(1+i)\Gamma(\frac{3-i}{2})^2}{\Gamma(1-i)\Gamma(\frac{3+i}{2})^2}~.
%
\end{align}
In order to implement our boundary condition at $u=1$, write $\mfh_{zz}$ in terms of $\Zone$:
\begin{align}
\mfh_{zz} = \frac{1+u^2}{u} \Zone'+\frac{2(1-u+u^2)}{1-u^2} \Zone~.
%
\end{align}
Then, near $u=1$, $\mfh_{zz}$ schematically behaves as
\begin{align}
\mfh_{zz} = c_1 \{ (u-1)^{1/2} +(u-1)^{3/2}+\cdots\}~,
%
\end{align}
where we ignore all numerical coefficients. But this does not satisfy our boundary condition $\mfh_{zz} ~\sim (u-1)$. Then, $c_1=0$, and there is no solution. 


\section{Discussion}\label{sec:discussion}

\begin{enumerate}
\item
When $\nq_z=0$, one obtains the mass spectrum $m_S<m_T<m_V$ like pure YM theories. 
In pure YM theories, the mass spectrum $m_S<m_T<m_V$ is qualitatively understood in terms of operator dimensions \cite{Kuti:1998rh}. The lowest dimension operator is dimension-4 operator $\text{tr}\calF_{ij}\calF_{kl}$, where $\calF_{ij}$ is the gauge field strength. One can form a spin-0 and spin-2 states, but one cannot form a spin-1 state. The lowest dimension operator with spin-1 comes from the dimension-5 operator $\text{tr}\calF_{ij}D_m\calF_{kl}$.

As pointed out in Ref.~\cite{Brower:2000rp}, the argument does not really apply to our holographic computation. We consider the gravitational perturbations, so in the dual gauge theory, we compute the spectrum of the energy-momentum tensor $T_{\mu\nu}$, which is a dimension-4 operator. One can form a spin-0, 1, and 2 states $T_{zz}, T_{zi}, T_{ij}$. Thus, the same pattern of the mass spectrum $m_S<m_T<m_V$  needs a different explanation. 

We argue that it is attributed to the pole-skipping. Namely, the mass spectrum $m_S<m_T<m_V$ for $\nq_z=0$ is just a coincidence. The appropriate order is $m_S<m_V<m_T$ by taking the pole-skipping into account. Then, the  relation is valid for all $\nq_z$. 

\item
The massless scalar field and the Maxwell scalar mode have the same spectrum when $\nq_z=0$. In fact, field equations are identical under an appropriate transformation.
They satisfy
\begin{subequations}
\begin{align}
0&=- \left(\frac{f}{u}\phi'\right)'+\frac{k^2}{u^2} \phi~,\\
0&= -\mfA_z''+ \frac{k^2}{uf} \mfA_z~.
\label{eq:maxwell_scalar0}
\end{align}
\end{subequations}
Setting $\tilA := f\phi'$ and taking the $u$-derivative of the scalar field equation gives
\begin{align}
0=-\tilA''+\frac{k^2}{uf}\tilA~,
%
\end{align}
which agrees with \eq{maxwell_scalar0}.
The boundary conditions are also identical. As $u\to0$,
\begin{align}
\phi\sim A+Bu^2~, \quad \mfA_z\sim A+Bu~.
%
\end{align}
Impose the source-free condition $A=0$ for the scalar. This implies $\tilA =f\phi' \sim Bu$, which reduces to the boundary condition for $\mfA_z$ ($A=0$). The $u\to1$ behavior is schematically written as
\begin{subequations}
\begin{align}
\phi &\sim c_1\{ 1+(u-1)+\cdots \}
\nonumber \\
&+c_2 \{\ln(u-1)+\cdots\}~, \\
\mfA_z &\sim c_1(u-1)\{1+\cdots\}+c_2\{1+\cdots\}~.
%
\end{align}
\end{subequations}
Impose the Neumann boundary condition $c_2=0$ for the scalar. This implies $\tilA =f\phi' \sim c_1(u-1)$, which reduces to the boundary condition for $\mfA_z$ ($c_2=0$). 
Because the field equation and the boundary condition for $\phi$ reduces to the ones for $\mfA_z$, they have the same spectrum. 

\item
The Maxwell scalar mode, gravitational vector and scalar modes have ``hydrodynamic" modes when $\nq_z,\nk\ll 1$ just like black hole cases. In the limit, one can find analytic solutions and can obtain dispersion-like relations which give mass spectrum (\appen{dispersion}). Of course, $\nq_z$ is actually discrete, so the dispersion relation gives only the approximate spectrum. 

\item 
Ref.~\cite{Brower:2000rp} computes normal modes for gravitational perturbations when $\nq_z=0$. Our results agree with Table~4 of Ref.~\cite{Brower:2000rp}. The relation between our convention and their convention is as follows: 
\begin{center}
\begin{tabular}{ccc}
Ours && Ref.~\cite{Brower:2000rp} \\
Massless scalar & $\leftrightarrow$ & $T_3 (0^{++})$ \\
Gravitational tensor & $\leftrightarrow$ & $T_3 (2^{++})$ \\
Gravitational vector & $\leftrightarrow$ & $V_3 (1^{++})$ \\
Gravitational scalar & $\leftrightarrow$ & $S_3 (0^{++})$ \\
\end{tabular}
\end{center}

\item
We study the pole-skipping in the AdS soliton geometry and argue that the pole-skipping has an interpretation as missing states. 
The analogous situation should occur in the \bh case. In the AdS soliton case, one cannot choose how one approaches a pole-skipping point in the $(\nq_z,\nk^2)$-plane. In the \bh case, one can choose how one approaches a pole-skipping point in the complex $(\omega,q)$-plane. But if one chooses a particular slope $\delta q/\delta \omega$, the pole-skipping point may not appear as a pole.

For example, near a pole-skipping point, the Green's function typically takes the form 
\begin{align}
G^R \propto \frac{\delta\omega+\delta q}{\delta\omega-\delta q}~,
%
\end{align}
so it depends on the slope $\delta q/\delta \omega$ how one approaches the point. However, if one first fixes $\delta q=0$, one gets $G^R =(\text{constant})$, so the pole disappears.
\end{enumerate}

\section*{Acknowledgments}


MN would like to thank Norikazu Yamada for useful discussions. 
This research was supported in part by a Grant-in-Aid for Scientific Research (17K05427) from the Ministry of Education, Culture, Sports, Science and Technology, Japan. 

\appendix

\section{Locations of regular singular points and alternative master variables}\label{sec:alternative}

\subsection{Alternative master variables}
We naively carry out the power-series expansion, but the power-series expansion is guaranteed to converge only inside a circle of radius $\rho$ around $u=1$ in the complex $u$-plane, where $\rho$ is the distance to the nearest singular point. Because we would like to extract the asymptotic behavior, $\rho$ should be greater than 1. However, our original master variables are problematic in this respect. This is pointed out, \eg, in Ref.~\cite{Kovtun:2005ev} in the context of the SAdS black hole, but it is rarely discussed in the literature. 

For the Maxwell scalar mode, the master equation \eqref{eq:master_diffusive} has singular points at 
\begin{align}
u=0, \pm1,\pm\sqrt{1+\nq_z^2/\nk^2}, \infty~. 
%
\end{align}
Note that there is a singular point whose location is $(\nq_z,\nk^2)$-dependent. The singular point appears in the form $1+\nq_z^2/\nk^2<1$, so the singular point lies in the region $0<u<1$,%
\footnote{The singular point may be pure imaginary when $1+\nq_z^2/\nk^2<0$.}
and the radius $\rho$ becomes less than 1. 
%
%
The power-series results indeed lie in this region. For the lowest $m_3^2$, the singular point lies at 
\begin{align}
(\nq_z,u)\sim (1,0.79)~, (2,0.67)~, (3,0.56)~, \cdots 
%
\end{align}
The gravitational vector mode \eqref{eq:master_shear} has the same singularity structure as the Maxwell scalar mode. 

For the gravitational scalar mode, the master equation \eqref{eq:master_sound} has singular points at 
\begin{align}
u=0, \pm1,\pm\sqrt{3(1+\nq_z^2/\nk^2)}, \infty~.
%
\end{align}
The radius $\rho$ becomes less than 1 when $1+\nq_z^2/\nk^2<1/3$. 
%
%
The power-series results indeed lie in this region.

This issue is not special to Frobenius method. Because the singular point lies in the region $0<u<1$, 
the shooting method also crosses the singular point which may be problematic. 

However, the choice of the master variable is not unique. One can choose an alternative master variable, where this problem does not occur. In principle, one should use such variables if one uses Frobenius method. However, it turns out that the results obtained by Frobenius method using an alternative master variable agree with the ones obtained by the naive Frobenius method using the original master variable.

\begin{enumerate}
\item
Maxwell scalar: we choose $\mfA_z$ as a master variable, but one can choose $\mfA_u$ as a master variable:
\begin{align}
0= Z_2''+\left( \frac{f'}{f}+\frac{1}{u} \right) Z_2'-\frac{\nq_z^2+\nk^2f}{uf^2}Z_2~,
%
\end{align}
where $Z_2:=f\mfA_u$. Asymptotically, $Z_2\sim A\ln u+B$. The master equation has singular points at $u=0, \pm1, \infty$, and one can safely use the power-series expansion.

For the master variable $\mfA_z$, one imposes the Dirichlet boundary condition at asymptotic infinity. One needs to determine the boundary condition for the master variable $Z_2$. From \eq{eom_Au}, the boundary condition $\mfA_z=0$ is translated as
\begin{align}
0=u(f\mfA_u)'=uZ_2'
%
\end{align}
for $Z_2$. Since $Z_2 \sim A \ln u+B$, the boundary condition reduces to $A=0$.

\item
Gravitational vector: we choose $\mfh_{zy}$ as a master variable, but one can choose $\mfh_{uy}$ as a master variable:
\begin{align}
0= Z_2''+\left( \frac{f'}{f}+\frac{2}{u} \right) Z_2'-\frac{\nq_z^2+\nk^2f}{uf^2}Z_2~,
%
\end{align}
where $Z_2:=f\mfh_{uy}$. Asymptotically, $Z_2\sim A u^{-1}+B$. 

For the master variable $\mfh_{zy}$, one imposes the Dirichlet boundary condition at asymptotic infinity. One needs to determine the boundary condition for the master variable $Z_2$. From \eq{eom_huy}, the boundary condition $\mfh_{uy}=0$ is translated as
\begin{align}
0=u(f\mfh_{uy})'=uZ_2'
%
\end{align}
for $Z_2$. Since $Z_2 \sim A u^{-1}+B$, the boundary condition reduces to $A=0$.
\end{enumerate}

\subsection{Middle-point prescription}
For the Maxwell scalar mode, the alternative master variable behaves as $Z_2\sim A\ln u+B$, 
and some care is necessary for numerical computations. First, impose the ansatz $Z_2=(1-u^2)^{\nq_z/2}\Za$. We carry out the power-series expansion both at $u=0$ and $u=1$. Denote the $u=1$ expansion which satisfies the boundary condition at $u=1$ as 
\begin{align}
\tilZ_A = 1+\cdots~, \quad(u\to1)~.
%
\end{align}
Denote the $u=0$ expansion which satisfies the boundary condition at $u=0$ as 
\begin{align}
\tilZ_B = b_0+\cdots~, \quad(u\to0)~.
%
\end{align}
We match $\tilZ_A, \tilZ_B$ at $u=u_0$ (\eg, $u_0=1/2$). Namely, demand that $\tilZ$ and $\tilZ'$ be continuous at $u=u_0$:
\begin{align}
\tilZ_{A}(u_0)=\tilZ_{B}(u_0)~, \quad
\tilZ_{A}(u_0)'=\tilZ_{B}(u_0)'~.
%
\end{align}
Because there are 2 constants $b_0$ and $k^2$, these conditions uniquely determine them.

Finally, alternative variables are suitable for normal mode analysis, but the original master variables are suitable for the boundary interpretation and for the pole-skipping analysis \cite{Natsuume:2023lzy}.

\subsection{Gravitational scalar}
For this mode, we are unable to find such an alternative master variable, and we use the shooting method. 
In the shooting method, one numerically integrates a differential equation. 
\begin{enumerate}
\item
Since the point $u=1$ is a regular singular point, one starts from $u=1-\epsilon$, where $\epsilon$ is a small number and impose a boundary condition there. 
\item
Then, one numerically integrates the equation from $u=1-\epsilon$ to $u=\epsilon$. (The point $u=0$ is a regular singular point as well.)
\item
In order to obtain normal modes, one imposes the source-free condition at $u=\epsilon$ by adjusting $m_3^2.$
\end{enumerate}

However, as pointed above, the issue of the singular point may be problematic for the shooting method as well.
Thus, we choose the integration path in the complex $u$-plane to avoid the singular point. We define
\begin{align}
u=\frac{1+e^{i\theta}}{2}~,
%
\end{align}
and integrate $\theta:0\to\pi$. Namely, we choose the integration path as the semicircle in the complex $u$-plane connecting $u=0$ and 1. The result agrees with the naive Frobenius method. 

\subsection{Interpretation}
All results agree with the ones by the naive Frobenius method using the original variable, so the singular point is not really problematic. We do not have a complete answer. But recall the construction of master equations. There are 2 gauge-invariant variables, and both obey first-order differential equations. These equations have singular points only at $u=0,\pm1,\infty$, and there is no $(\nq_z,\nk^2)$-dependent singular point. This additional singular point appears when one combines these 2 equations to obtain master equations. But one could apply Frobenius method in the coupled equations, where this issue does not arise. Probably, one needs to take into account only the singular points which appear in coupled equations.

\section{``Hydrodynamic" limit}\label{sec:dispersion}

\begin{table}[tb]
\begin{center}
\begin{tabular}{l|r|cc}
&& ``hydrodynamic" & numerical \\
\hline
Maxwell scalar	
				& $\nq_z=1$ & 4 & 7.162  \\
				& $=2$ & 8 & 21.45  \\
Gravitational vector
				& $\nq_z=1$ & 8 & 10.79  \\
				& $=2$ & 16 & 28.04  \\
Gravitational  scalar
				& $\nq_z=1$ & 12 & (6.000)  \\
				& $=2$ & 48 & 20.27 \\
%
\end{tabular}
\caption{The lowest $m_3^2$ obtained by the extrapolation of the hydrodynamic dispersion relation and by  numerical computations.}
\label{fig:hydro}
\end{center}
\end{table}

For the SAdS$_5$ black hole, hydrodynamic poles $\nw, \nq\ll1$ appear in the Maxwell scalar mode, gravitational vector and scalar modes. Because the master equations for the AdS soliton are obtained from the \bh by the double Wick rotation, one expects ``hydrodynamic" modes when $\nq_z,\nk\ll 1$. Of course, $\nq_z$ is actually discrete, so there is really no hydrodynamic mode. 

In the hydrodynamic limit $\nk\ll1$ and $\nq_z\ll1$, one can find analytic solutions.
Set $\nk\to \epsilon \nk, \nq_{z}\to \epsilon \nq_{z} $ and expand the master variables $\Za$ as a series in $\epsilon$:
\begin{align}
\Za=Z^{(0)}+\epsilon Z^{(1)}+\cdots~.
%
\end{align}
We impose the boundary conditions (1) regular at $u=1$ [and $Z^{(1)}(1)=0$] and (2) the source-free condition at $u=0$. 

For the Maxwell scalar mode, the solution is given by ($\epsilon\to1$)
\begin{align}
\Za &= C_0 \left[1+\frac{\nk^2(1-u)}{\nq_z} -\nq_z\ln\frac{1+u}{2} +\cdots \right]~.
%
\end{align}
The asymptotic behavior is given by
\begin{align}
\Za \sim \frac{C_0}{\nq_z}(\nq_z+\nk^2+\nq_z^2\ln 2)+O(u)~,
\quad(u\to0)~.
%
\end{align}
The source-free condition then gives a ``dispersion relation"
\begin{align}
\nq_z=-\nk^2 +\cdots~.
%
\end{align}
This corresponds to the diffusion pole in the SAdS$_5$ case:
\begin{align}
\nw = -i\nq^2+\cdots~,
%
\end{align}
where $\nw:=\omega/(2\pi T), \nq:=q/(2\pi T)$. To obtain the spectrum, rewrite the dispersion relation in terms of mass:
\begin{align}
m_3^2:=-p^2=4\nq_z+\cdots~.
%
\end{align}

The dispersion relation for the AdS soliton should not be taken literally because $\nq_z$ is actually discrete. One should regard the dispersion relation as an approximate relation of the spectrum given an integer $\nq_z$. For example, the dispersion relation gives $m_3^2\approx4$ for $\nq_z=1$. On the other hand, the computation in \sect{spectrum} gives $m_3^2\approx 7.16$. Because $\nq_z$ is $O(1)$, the dispersion relation does not give a very reliable result. 
See Table~\ref{fig:hydro} for the other modes.

For the gravitational vector mode, the solution is given by
\begin{subequations}
\begin{align}
\Za 
&= C_0 \left[ 1 +\frac{\nk^2}{2\nq_z}(1-u^2)+\cdots \right] \\
&\sim C_0 \frac{2\nq_z + \nk^2}{2\nq_z} + O(u^2)~, 
\quad (u\to0)~.
%
\end{align}
\end{subequations}
Then, the dispersion relation is
\begin{align}
\nq_z &= -\frac{1}{2}\nk^2 +\cdots~,
%
\end{align}
or $m_3^2=-p^2=8\nq_z+\cdots$. This corresponds to the shear pole in the SAdS$_5$ case: $\nw=-i\nq^2/2+\cdots$.

For the gravitational scalar mode, the solution is given by
\begin{subequations}
\begin{align}
\Za &= C_0 \left[ \frac{3\nq_z^2+\nk^2\{ 1+u^2+2\nq_z(1-u^2) \}}{3\nq_z^2+2\nk^2}+\cdots \right]~, \\
&\sim C_0 \frac{3\nq_z^2+\nk^2+2\nq_z\nk^2}{3\nq_z^2+2\nk^2} +O(u^2)~, 
\quad (u\to0)~,
%
\end{align}
\end{subequations}
and the dispersion relation is
\begin{subequations}
\begin{align}
0 &\sim 3\nq_z^2+\nk^2+2\nq_z\nk^2~, \\
\to \nq_z &=\pm \sqrt{ \frac{-\nk^2}{3}}  - \frac{1}{3}\nk^2 +\cdots~,
%
\end{align}
\end{subequations}
or $m_3^2=12\nq_z^2+\cdots$.
This corresponds to the sound pole in the SAdS$_5$ case: $\nw = \pm\nq/\sqrt{3}-i\nq^2/3+\cdots$.


\end{document}